\documentclass[preprint,showpacs,preprintnumbers,amsmath,amssymb]{revtex4}

\usepackage{graphicx}
\usepackage{dcolumn}
\usepackage{bm}
\usepackage{natbib}
\usepackage{braket}
\usepackage{units}

\renewcommand{\d}{ \text{d}}

\begin{document}

\preprint{DO-TH-09/03}

\title{Coherent Pion Production by Neutrinos}

\author{E. A. Paschos}
 \email{paschos@physik.uni-dortmund.de}
\author{Dario Schalla}
 \email{dario.schalla@tu-dortmund.de}
\affiliation{Department of Physics, TU Dortmund, D-44221 Dortmund, Germany}

\date{May 7, 2009}

\begin{abstract}
In this paper we present quantitative results for coherent pion production by neutrinos scattered off nuclei within the framework developed by Gounaris, Kartavtsev and Paschos. The method is based on PCAC and uses helicity cross sections for the scattering of weak gauge bosons on nuclei. The process relies on experimental data for elastic pion-nucleus scattering. A detailed analysis of the differential and integrated cross sections is presented for neutral and charged currents, with special emphasis on the regions of integrations. The results are extended to energies of $10.0 \unit{GeV}$ and are compared with experimental data.
\end{abstract}


\maketitle

\section{Introduction}
\label{sec:Introduction}
The process of coherent pion production by neutrinos is one of the important reactions that occur in neutrino oscillation experiments. It dominates at very low momentum transfers of the leptons and in the forward direction which may help to monitor the flux of the beam and will help to estimate backgrounds.  For instance in the oscillation of $\nu_\mu$  into $\nu_e$ one searches in the far away detector the reaction $\nu_e N \rightarrow e^- X$ and coherent production of $\pi^0$'s produces a background. For this reason there are new theoretical articles and experimental mesurements dealing with this process.

The neutral and charged current reactions are
\begin{eqnarray}
\nu_\mu (k) N (p) &\rightarrow& \nu_\mu (k') N (p') \pi^0 (p_\pi) \\
\nu_\mu (k) N (p) &\rightarrow& \mu^- (k') N (p') \pi^+ (p_\pi) 
,\end{eqnarray}
where $N$ is a nucleus. 
For the process we use variables in the rest frame of the nucleus with $q=k-k'$, $Q^2 = - q^2$, $\nu=E-E'$ and $t=(q-p_\pi)^2$.
Coherent production implies that the nucleus does not break up or alter its quantum numbers during the process.

\begin{figure}
\includegraphics[width=6cm]{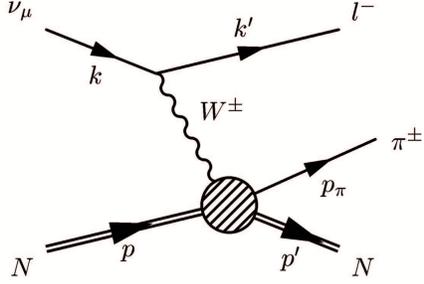}
\caption{Feynman diagram of the charged current reaction}
\label{fig:CCFeyDia}
\end{figure}

Due to the isospin structure of the charged and neutral currents, the charged cross section is at high energies, approximately, twice as big as the neutral one. At low energies difference arise from the non-vanishing mass of the outgoing lepton and the Cabibbo-Kabayashi-Maskawa-matrix element (CKM) in the charged current case. Hence it is sufficient to concentrate on the charged current reaction within the theory. Its Feynman diagram is shown in figure \ref{fig:CCFeyDia}.

Our calculation \cite{Paschos:2005km} is based on two basic facts:\\
1. Coherent production of pions is the process where the four-momentum-transfer squared between the current and the produced pion is small so the nucleus remains intact. The minimum value is given to a good approximation by
\begin{equation}
|t_{min}| = \left( \frac{Q^2+m_\pi^2}{2 \nu} \right)^2
\end{equation}
and is achieved for $\nu M >> Q^2$. The important experimental region will turn out to be $Q^2< 0.1 \unit{GeV^2}$. We remind the reader that, neglecting the nucleus recoil energy, $t$ is calculated as \cite{Allport:1988cq}
\begin{equation}
|t| = \left[ \sum_{\mu, \pi} p_i^\perp \right]^2 + \left[ \sum_{\mu, \pi} \left( E_i - p_i^\parallel \right) \right]^2
.\end{equation}

2. In this region, the dominant component of the leptonic current has helicity zero, i.e. $\epsilon_\mu (\lambda = 0 )$ defined in equation (\ref{lambda0}), below. The domain $\nu M >> Q^2$ and $Q^2 = \text{(a few) } m_\pi^2$ is the region where PCAC is valid and it has been shown that the dominant amplitude is determined by chiral symmetry. In fact when we write the amplitude for the subprocess $WN\rightarrow N\pi$ as the sum
$$  \text{pion pole } + \mathcal{R}_\mu $$
with $\mathcal{R}_\mu$ the remainder, the symmetry determines
\begin{equation} q_\mu \mathcal{R}^\mu = - f_\pi \sqrt{2} T (\pi^+ N \rightarrow \pi^+ N ) \end{equation}
which is exactly the term that the zero helicity polarization selects and makes it dominant. In this picture the weak current converts into a pion before entering the nucleus. We shall quantify these results in the following sections.\\

This approach has many advantages:
\begin{enumerate}
\item It avoids discussing what happens within the nucleus since it will use $\pi N$ elastic scattering data. In this sense it incorporates the pion-nucleus diffrative peak into the neutrino scattering.
\item It is applicable at low and high energies, provided that some required kinematic cuts are made.
\item It includes the lepton mass exactly and quantifies the various approximations. 
\end{enumerate}

The purpose of this article is to quantify the results with numerical calculations. There are other articles on this topic \cite{Berger:2008xs,AlvarezRuso:2007tt,AlvarezRuso:2007it,Singh:2006bm,Amaro:2008hd,Nakamura:2009gg}. One among them follows the same approach \cite{Berger:2008xs}. A very different approach descibes coherent pion production at very low energies as the excitation of the $\Delta$-resonance and includes modifications through nuclear medium effects demanding that the nucleus remains in its ground state for coherent scattering \cite{AlvarezRuso:2007tt,AlvarezRuso:2007it,Singh:2006bm,Amaro:2008hd,Nakamura:2009gg}.

\section{Theory Revisited}
\label{sec:TheoryRevisited}
The theoretical background of this paper is the same as in \cite{Paschos:2005km}. For completeness we repeat its main features. The invariant matrix element of the charged current reaction is
\begin{equation} \mathcal{M} = - \frac{G_F V_{ud}}{\sqrt{2}} j_\mu  \Braket{\pi N | J^\mu | N} ,\end{equation}
where $G_F$ is the Fermi coupling constant and $V_{ud}$ the CKM matrix element. The leptonic current is expressed in a straight forward way
\begin{equation} j_\mu = \overline{u} (k') \gamma_\mu \left( 1 - \gamma_5 \right) u(k) \end{equation}
whereas the hadronic matrix element will be treated below. 

The leptonic current is decomposed into the basis of the four polarisation vectors of the exchange boson. Select the 3-momentum of $q_\mu$ along the $z$-axis and define the basis of polarization vectors
\begin{equation} \epsilon^l_\mu = \frac{q_\mu}{\sqrt{Q^2}} = \frac{1}{\sqrt{Q^2}} \begin{pmatrix} \nu \\ 0 \\ 0 \\ |\vec{q}| \end{pmatrix} \end{equation}
\begin{equation} \epsilon_\mu (\lambda \pm 1 ) =  \frac{1}{\sqrt{2}} \begin{pmatrix} 0 \\ 1 \\ \pm i \\ 0 \end{pmatrix} , \; \epsilon_\mu (\lambda=0) = \frac{1}{\sqrt{Q^2}} \begin{pmatrix} |\vec{q}| \\ 0 \\ 0 \\ \nu \end{pmatrix} . \label{lambda0} \end{equation}
Note that they differ from the original paper  \cite{Paschos:2005km} by their normalization factors. This has been taken into account in the following formulas but does not change the results. The polarisation vectors satisfy the completeness relation
\begin{equation}
\sum_{\lambda = 0,\pm 1} (-1)^\lambda \epsilon^\mu (\lambda) \epsilon^{\nu *} (\lambda) - \epsilon_l^\mu \epsilon_l^\nu = g^{\mu\nu} 
. \label{compl} \end{equation}

In the cross sections the leptonic tensor appears
\begin{eqnarray}
T_{\mu\nu} &=& 8 (k_\mu k_\nu' + k_\nu k_\mu' - g_{\mu\nu} k\cdot k' ) + 8i \varepsilon_{\mu\alpha\nu\beta} k^\alpha k'^\beta \\
&=& \sum_{JJ'} L_{JJ'} \epsilon_\mu (J) \epsilon_\nu^* (J')
\label{leptensor}\end{eqnarray}
with the indices running over the four polarisations. Numerical studies in reference \cite{Paschos:2005km} have shown that the transverse polarizations, i.e. right and left handed cross sections are smaller in coherent scattering and they will be omitted.

From equation (\ref{leptensor}) we obtain the relevant matrix elements:
\begin{eqnarray}
L_{00} &=& 4 \frac{\left[ Q^2(2E_\nu - \nu) - \nu m_\mu^2 \right]^2}{Q^2 (Q^2 + \nu^2) } - 4 (Q^2 + m_\mu^2)\label{L00}\\
L_{l0} &=& 4 m_\mu^2 \frac{Q^2 (2E_\nu - \nu ) - \nu m_\mu^2}{ Q^2 \sqrt{Q^2+\nu^2}} \\
L_{ll} &=& 4 m_\mu^2 \left( 1 + \frac{m_\mu^2}{Q^2} \right)
.\end{eqnarray}

Here the mass of the lepton in the final state is explicit and will be kept throughout the calculation.
We also use later $\tilde{L}_{ij}$ as in reference \cite{Paschos:2005km} and the relation between them is $\tilde{L}_{ij} = \frac{1}{2} L_{ij}$.

It is now possible to write the spin averaged matrix element as
\begin{widetext}
\begin{equation}
\overline{|\mathcal{M}|^2} = \frac{G_F^2 |V_{ud}|^2}{2} \left\{ L_{00} |J_\mu \epsilon_0^\mu |^2 + L_{ll} |J_\mu \epsilon_l^\mu |^2 \right. \\
+ \left. 2 L_{l0} ( J_\mu \epsilon_l^\mu ) ( J_\mu\epsilon^\mu_0 )^* \right\}
.\label{msq} \end{equation}
\end{widetext}

As mentioned in the introduction the divergence of the matrix element is determined by PCAC. The matrix element of the axial current can be written as a sum of the pion pole and the remaining contribution
\begin{equation}
-i\mathcal{A}^+_\mu = \frac{\sqrt{2} f_\pi q_\mu}{Q^2+m_\pi^2} T(\pi^+ N \rightarrow \pi^+ N) - \mathcal{R}_\mu
\end{equation}
where $T(\pi^+ N \rightarrow \pi^+ N)$ is the elastic scattering on a nucleus and $\mathcal{R}_\mu$ is a smooth function which includes all other contributions. The PCAC relation gives
\begin{equation}
-i q^\mu \mathcal{A}_\mu^+ = \frac{\sqrt{2} f_\pi m_\pi^2}{Q^2+m_\pi^2} T(\pi^+ N \rightarrow \pi^+ N)
.\label{qA}\end{equation}
Combining the two equations one obtains
\begin{equation}
q^\mu \mathcal{R}_\mu = - \sqrt{2} f_\pi T(\pi^+ N \rightarrow \pi^+ N )
.\end{equation}

With these relations one is also able to calculate all matrix elements in equation (\ref{msq}) because
\begin{equation}
\epsilon^l_\mu \mathcal{A}^{\mu +} = -i \frac{q^\mu A_\mu^+}{\sqrt{Q^2}}
\end{equation}
is given by equation (\ref{qA}) and is proportional to $m_\pi^2$.

The matrix element for helicity zero is calculated in the following way. From the property $\epsilon^\mu (\lambda=0) q_\mu = 0$ it follows that the inner product of the pion pole with the helicity zero polarization vanishes. The remaining term is now estimated
\begin{equation}
\epsilon^\mu (\lambda =0) \mathcal{R}_\mu = \frac{q^\mu}{\sqrt{Q^2}} \mathcal{R}_\mu + \mathcal{O} \left( \frac{\sqrt{Q^2}}{\nu} \right) \\
=  -\frac{\sqrt{2} f_\pi}{\sqrt{Q^2}} T (\pi^+ N \rightarrow \pi^+ N) + \mathcal{O} \left( \frac{\sqrt{Q^2}}{\nu} \right) \label{remainingterm}
.\end{equation}
Thus the matrix elements for both polarizations are estimated.

We write next the triple differential cross section as
\begin{widetext}
\begin{equation}
\frac{\d \sigma_{CC}}{\d Q^2 \d \nu \d t} = \frac{G_F^2 |V_{ud}|^2}{ 2 (2\pi)^2} \frac{\nu}{E_\nu^2} \frac{f_\pi^2}{Q^2} \left\{ \tilde{L}_{00} + \tilde{L}_{ll} \left( \frac{m_\pi^2}{Q^2+m_\pi^2} \right)^2 + 2 \tilde{L}_{l0} \frac{m_\pi^2}{Q^2+m_\pi^2}  \right\} \frac{\d \sigma_\pi}{\d t}
,\label{CC} \end{equation} \end{widetext}
where $\sigma_\pi$ is the elastic pion nucleus cross section and where all muon mass terms have been kept in the calculation. As mentioned already the right- and left-handed cross sections are small and have been neglected. Since they appear as positive additive terms, our estimate is a lower bound for the coherent cross section.

The corresponding neutral current cross section is
\begin{equation}
\frac{\d \sigma_{NC}}{\d Q^2 \d \nu \d t} = \frac{G_F^2}{4 (2 \pi)^2} \frac{\nu}{E_\nu^2} \frac{f_\pi^2}{Q^2} \tilde{L}_{00} \frac{\d \sigma_\pi}{\d t}
. \label{NC} \end{equation} 
In this formula the muon mass in $\tilde{L}_{00}$ has to be set zero.

\section{Methods of Integration}
\label{sec:Kinematics}
The most detailed and convincing evidence for coherent scattering is the explicit observation of the triple differential cross sections in equations (\ref{CC}) and (\ref{NC}). The characteristic signature is the sharp peak in the $t$-distributions which was the main feature in the original discovery and interpretation \cite{Faissner:1983ng,Rein:1982pf}. Subsequent and recent experiments integrate over the $t$- and other variables so that special attention must be given to the ranges of integration in order to ascertain that the model is still valid in these regions. In this article we shall integrate over data for $\pi^+ \textrm{C}^{12}$ elastic scattering.

We identify the incoming pion energy with the variable $\nu$ and integrate over experimental data for elastic pion-nucleus scattering \cite{Edelstein,Binon:1970ye,Takahashi:1995hs}. We parametrize the cross section as follows:
\begin{equation}
\frac{\d \sigma_\pi}{\d t} = a \exp [ -b |t| ]
\end{equation}
and fitted the parameters $a$ and $b$ to the data. Their numerical values are given in table \ref{tab:model}.

\begin{table}[h]
\begin{tabular}{|c|c|c|}
\hline 
$\nu$ [GeV]		&	$a$ [barn/$\text{GeV}^2$]	&	b [$\text{GeV}^{-2}$]	\\
\hline
0.210	&	28.526	&	159.657	\\
0.228	&	28.659	&	147.986 \\
0.260	&	32.012	&	129.022	\\
0.290	&	27.162	&	101.910	\\
0.320	&	23.600	&	90.824	\\
0.340	&	22.734	&	90.660	\\
0.370	&	19.000	&	83.814	\\
0.400	&	17.924	&	84.590	\\
0.420	&	14.594	&	73.256	\\
0.766	&	3.759	&	49.459	\\
0.864	&	4.172	&	58.149	\\
0.942	&	3.649	&	56.197	\\
1.046	&	3.523	&	53.497	\\
\hline
\end{tabular}
\caption{Parameters of the elastic pion nuclues cross section model.}
\label{tab:model}
\end{table}

As in reference \cite{Paschos:2005km} we integrate first the variable $t$ over the range
\begin{equation}
\left( \frac{Q^2 + m_\pi^2}{2\nu} \right)^2 \leq |t| \leq \infty
.\label{trange}\end{equation}
The upper limit of integration should be the first diffractive minimum and has been expanded to infinity because the numerical results are insensitive to values of this cross sections beyond the first diffractive minimum, the cross section being already too small. The other limit of integration is important at low energies and influences the $Q^2$ dependence. The integrated elastic pion-nucleus cross section
\begin{equation}
\sigma_\pi (Q^2,\nu) = \int_{t_{min}}^\infty \frac{\d \sigma_\pi}{\d t} \d t
\end{equation}
depends on $Q^2$ and $\nu$ and the results are shown in figure 2. In reference \cite{Paschos:2005km} the $\nu$-range was limited to $0.4\unit{GeV}$ and now we extended it to $\nu = 1.0\unit{GeV}$. The data has been extrapolated as constants in regions not covered by data. We observe again a sharp $Q^2$-dependence.\\

\begin{figure}
\includegraphics[scale=1.5]{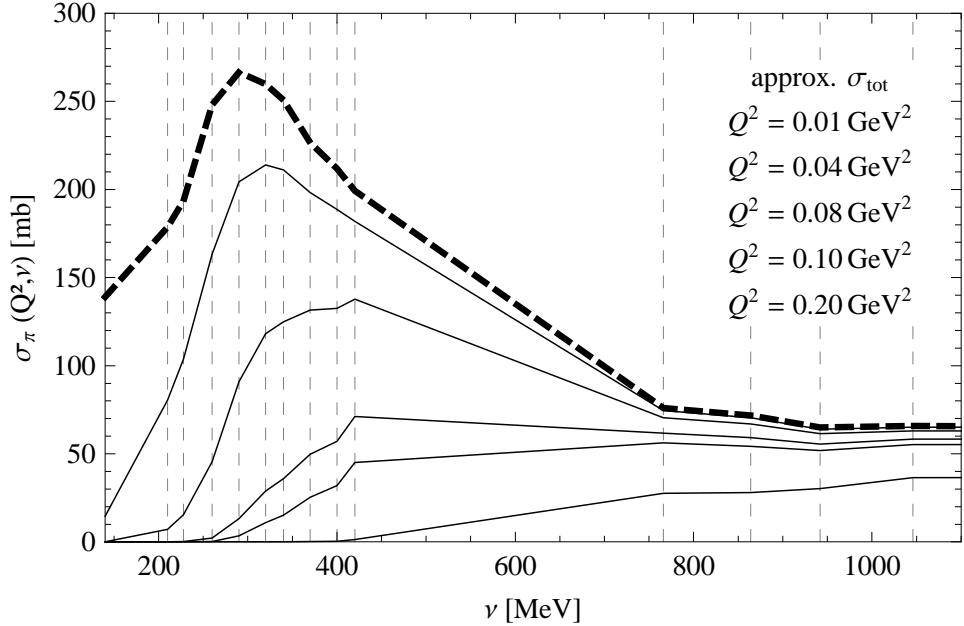}
\caption{Integrated elastic pion nucleus cross section for some momentum transfers.}
\label{fig:hadron}
\end{figure}

\begin{figure}[h]
\includegraphics[scale=1]{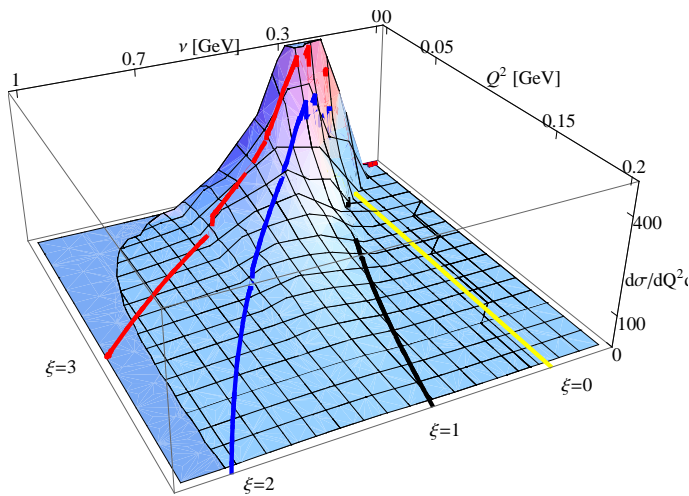}
\caption{Differential charged current cross section at $E_\nu=1\unit{GeV}$. The lines represent different integration limits (see text).}
\label{fig:dE1cc}
\end{figure}

\begin{figure}[h]
\includegraphics[scale=1]{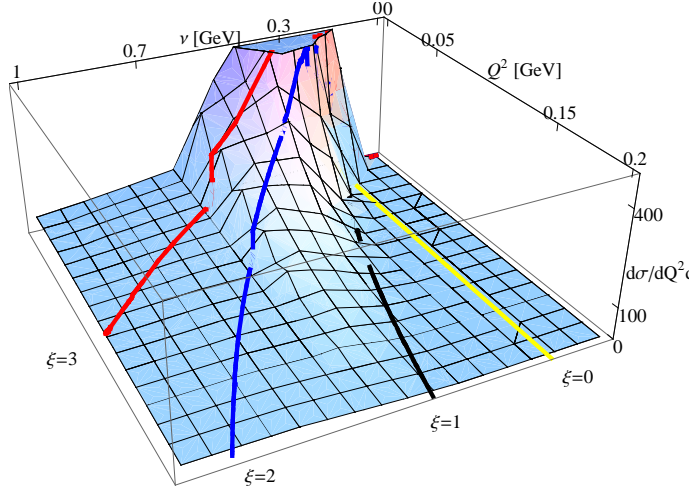}
\caption{Differential charged current cross section at $E_\nu=10\unit{GeV}$. The lines represent different integration limits (see text).}
\label{fig:dE10cc}
\end{figure}

The integration over $\nu$ should respect the condition
\begin{equation}
\nu >> \sqrt{Q^2}
\end{equation}
so that the helicity $\lambda =0$ polarization can be expanded to give
\begin{equation}
\epsilon_\mu (\lambda = 0) \approx \frac{q_\mu}{\sqrt{Q^2}} + \mathcal{O} \left( \frac{\sqrt{Q^2}}{\nu} \right)
\end{equation}
which has been used in equations (\ref{remainingterm} - \ref{NC}). In order to test this condition we computed the double differential cross section $\frac{\d \sigma}{\d Q^2 \d \nu}$ over the entire range of $Q^2$ and $\nu$ using as input the curves in figure \ref{fig:hadron}. The results are shown in figures \ref{fig:dE1cc} and \ref{fig:dE10cc} where we also include four curves. These four curves correspond to the kinematic regions defined in table \ref{tab:PhaseSpace}.

\begin{table}
	\centering
		\begin{tabular}{|c|c|}
			\hline
				condition	&	value of $\xi$ in reference [1]	\\
			\hline
				$\nu = \nu_{min}$		&	$\xi=0$	\\
				$\nu = \sqrt{Q^2}$	&	$\xi=1$	\\
				$\nu = 2\sqrt{Q^2}$	&	$\xi=2$	\\
				$\nu = 3\sqrt{Q^2}$	&	$\xi=3$	\\
			\hline
		\end{tabular}
	\caption{Defining regions of phase space.}
	\label{tab:PhaseSpace}
\end{table}

In the earlier article \cite{Paschos:2005km} we used the variable $\xi$, which defined kinematic regions available in the experiments. It is a fortunate property of coherent scattering that the cross section peaks at low values of $Q^2$ and $\nu$. In fact the contribution to the integrated cross section from the region
\begin{equation} 0 < \nu < \sqrt{Q^2} \end{equation}
is negligible. Our estimates are satisfied for $\nu > 1.5 \sqrt{Q^2}$ where most of the cross section is located. One can also integrate over restricted regions of the phase space in order to determine the fraction of the cross section in these regions.

The structure of the cross section in figure \ref{fig:dE1cc} also indicates the important regions of integration. To assure the validity of the approximation in equation (\ref{remainingterm})  we integrate over the range 
\begin{equation}
\max  \left( \xi \sqrt{Q^2} , \nu_{min} \right) < \nu < \nu_{max}
.\end{equation}
The values for $\nu_{min}$ and $\nu_{max}$ are given in the appendix of reference \cite{Paschos:2005km}.

The $Q^2$ integration is treated more carefully. The lower limit of integration is given in equation (A8) of reference \cite{Paschos:2005km}. In the neutral current case it is zero.

\begin{figure}[h]
\includegraphics[scale=1.4]{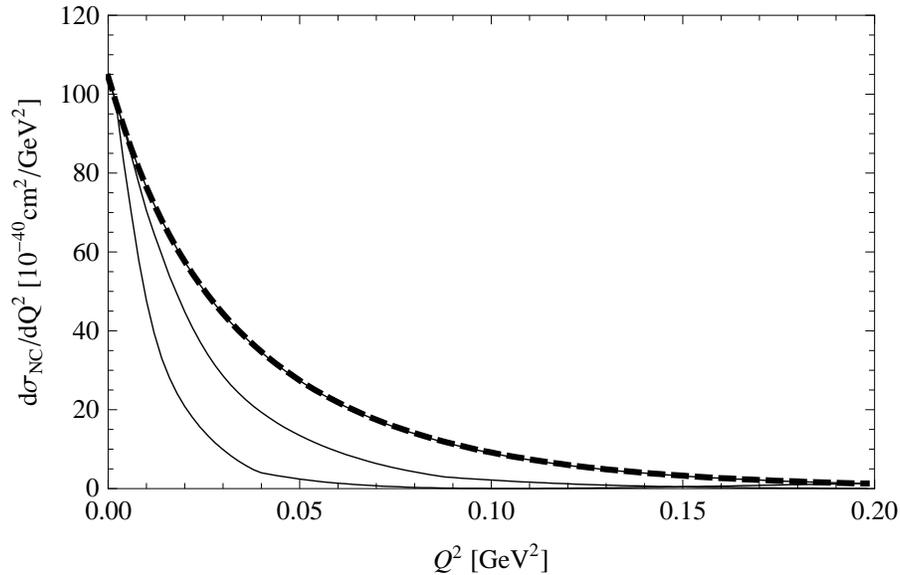}
\caption{Differential neutral current cross section for $\xi=1,2,3$ (top to bottom) and $\xi=0$ (dashed) at $E_\nu=1\unit{GeV}$.}
\label{fig:e1nallxi}
\end{figure}

\begin{figure}[h]
\includegraphics[scale=1.4]{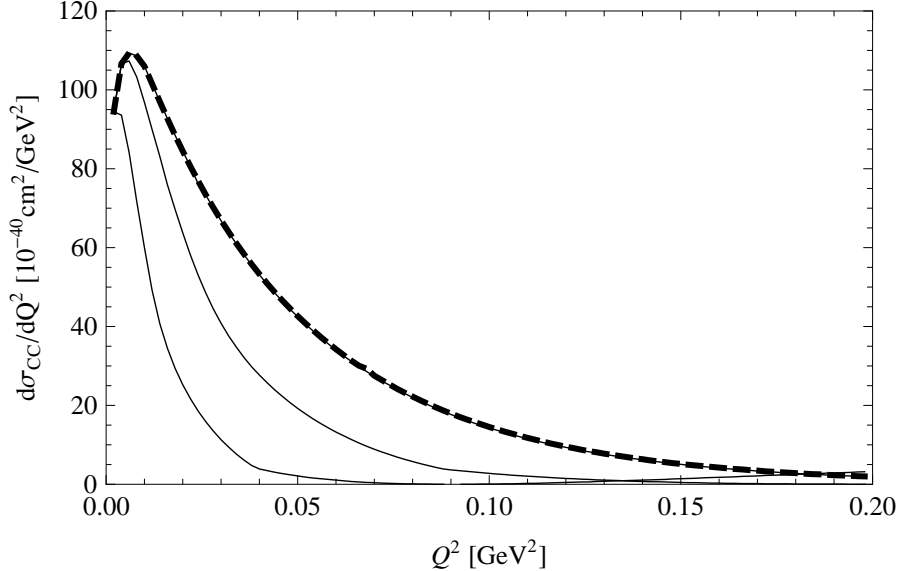}
\caption{Differential charged current cross section for $\xi=1,2,3$ (top to bottom) and $\xi=0$ (dashed) at $E_\nu=1\unit{GeV}$.}
\label{fig:e1callxi}
\end{figure}

We calculated the differential cross section $\frac{\d \sigma}{\d Q^2}$ for $E_\nu = 1\unit{GeV}$ and four values for the lower limit of the energy $\nu$. The results are shown for neutral currents in figure \ref{fig:e1nallxi} and for charged currents in figure \ref{fig:e1callxi}. The four curves correspond to various cuts in the minimum value of the energy $\nu$, defined in table \ref{tab:PhaseSpace}. We note that the cross section between $\nu_{min}<\nu<\sqrt{Q^2}$ is negligibly small since the dashed curves for $\nu > \nu_{min}$ coincides with curve for $\nu > \sqrt{Q^2}$.

\begin{figure}[h]
\includegraphics[scale=1.5]{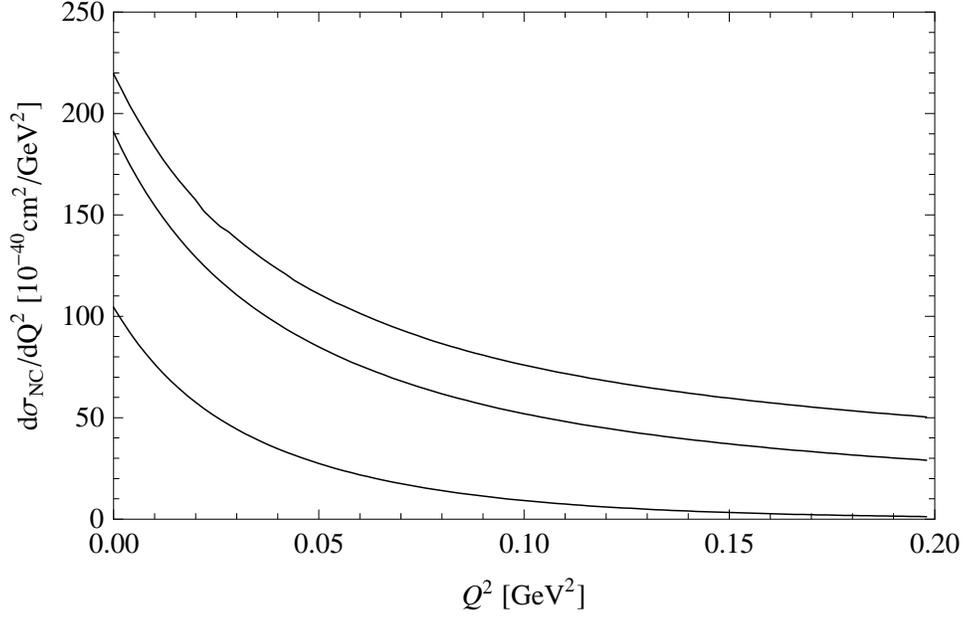}
\caption{Differential neutral current cross section for $\xi=0$ at $E_\nu=1,5\text{ and }10\unit{GeV}$.}
\label{fig:e1051nxi0}
\end{figure}

\begin{figure}[h]
\includegraphics[scale=1.4]{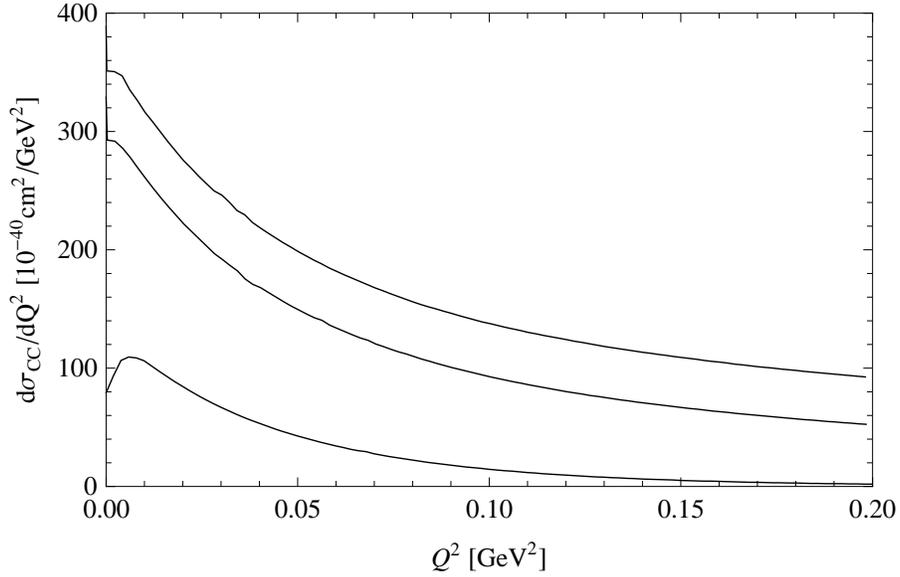}
\caption{Differential charged current cross section for $\xi=0$ at $E_\nu=1,5\text{ and }10\unit{GeV}$.}
\label{fig:e1051cxi0}
\end{figure}

We repeated this calculation for various incident neutrino energies and the results are shown in figures \ref{fig:e1051nxi0} and \ref{fig:e1051cxi0} for neutral and charged currents, respectively. The interesting feature is that the cross section extends now to higher values of $Q^2$. Although the differential cross section is concentrated at low momentum transfers, for higher neutrino energies there is a tail that extends to larger values of $Q^2$ and must be taken into account in the integrated cross sections.

The neutral current cross section reaches a specific value at $Q^2=0$, the Adler point. It is estimated from our formulas (\ref{L00}) and (\ref{NC}) to give
\begin{eqnarray}
\left. \frac{\d \sigma_{NC}}{\d Q^2} \right|_{Q^2=0} &=& \frac{G_F^2f_\pi^2}{2\pi^2} \int_{m_\pi}^{E_\nu} \d \nu \; \sigma_\pi (\nu) \left\{ \frac{1}{\nu} - \frac{1}{E_\nu} \right\} \\
&=& \frac{G_F^2 f_\pi^2}{2\pi^2} <\sigma_\pi> \left\{ \ln \frac{E_\nu}{m_\pi} + \frac{m_\pi}{E_\nu} - 1 \right\}
\end{eqnarray}
where $<\sigma_\pi>$ is the weighted elastic pion-nucleus cross section. For $E_\nu=1\unit{GeV}$  the Adler point is about $112\cdot10^{-40}\unit{\frac{cm^2}{GeV^2}}$ for $<\sigma_\pi> \approx 165\unit{mb}$. This is a realistic average of the weighted hadronic cross section (see figure \ref{fig:hadron}). A similar comparison for the charged current is more indirect because of the phase space effects introduced by the mass of the muon. For the charged current an integration over a small region of $Q^2$ is more appropriate for comparison with experiments.

It is worth mentioning that for higher energies the turnover of the differential cross section in the charged current case is less prominent and disappears, a property caused by the fact that the muon mass is negligible compared to the neutrino energy.

At higher neutrino energies bigger momentum transfers gain more importance. Figures \ref{fig:e1051nxi0} and \ref{fig:e1051cxi0} show the differential cross section of the neutral and charged current process, respectively, for $\xi=0$. For higher energies the cross section is not limited within a small $Q^2$ region. Thus the $Q^2$ integration interval has to be extended to higher values.

\section{Integrated Cross Section}
\label{sec:IntegratdCrossSection}
The results in figure \ref{fig:e1051cxi0} indicate that the formulas we derived have a tail that extends to higher values of $Q^2$. However, at large values of $Q^2$ the nucleus breaks up and the scattering is not coherent. At the same time the PCAC approximation does not hold. Fortunately, the change in the curves of figures 9 and 20 between $\xi=1$ and 4 is relatively small. Experimental groups separate the coherent events from other events by the absence of stubs in bubble chambers, a signature indicating that the nucleus does not break up. The BEBC group \cite{Marage:1986cy} found that the $Q^2$-values are lower for coherent events and do not extend beyond $2.0\unit{GeV^2}$. These results suggest that a $Q^2_{max}$ must be introduced in the integration of the cross section or a phenomenological factor $\left( \frac{m_A^2}{Q^2 + m_A^2} \right)^2$ in order to provide a cut-off in the $Q^2$ dependence (see \cite{Belkov:1986hn}).

\begin{figure}[h]
\includegraphics[scale=1.4]{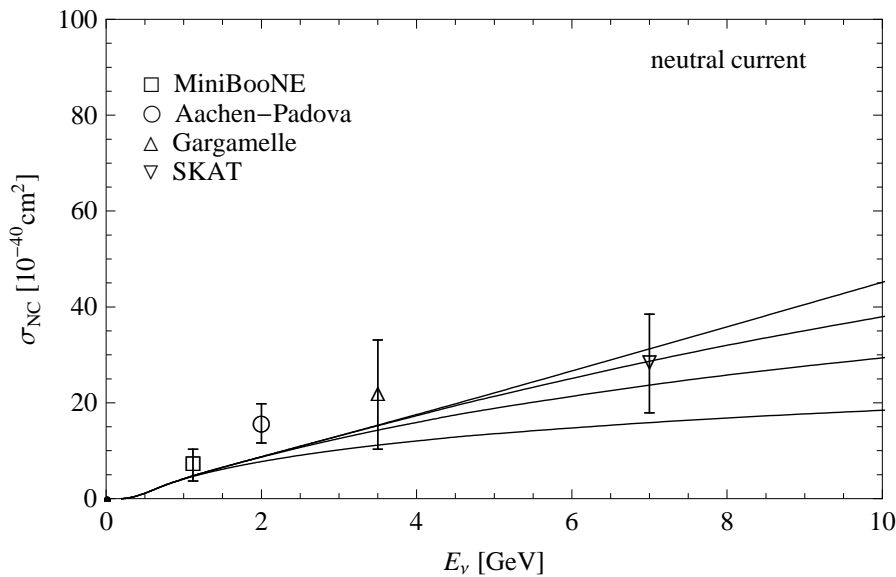}
\caption{Integrated neutral current cross section with $Q^2_{max}=0.2, \; 0.5, \; 1.0 \text{ and } 4.0\unit{GeV^2}$ for $\xi=0$ (bottom to top). See \cite{Raaf:2005up,Faissner:1983ng,Isiksal:1984vh,Grabosch:1985mt} for experimental data.}
\label{fig:totalNCcuts}
\end{figure}

\begin{figure}[h]
\includegraphics[scale=1.4]{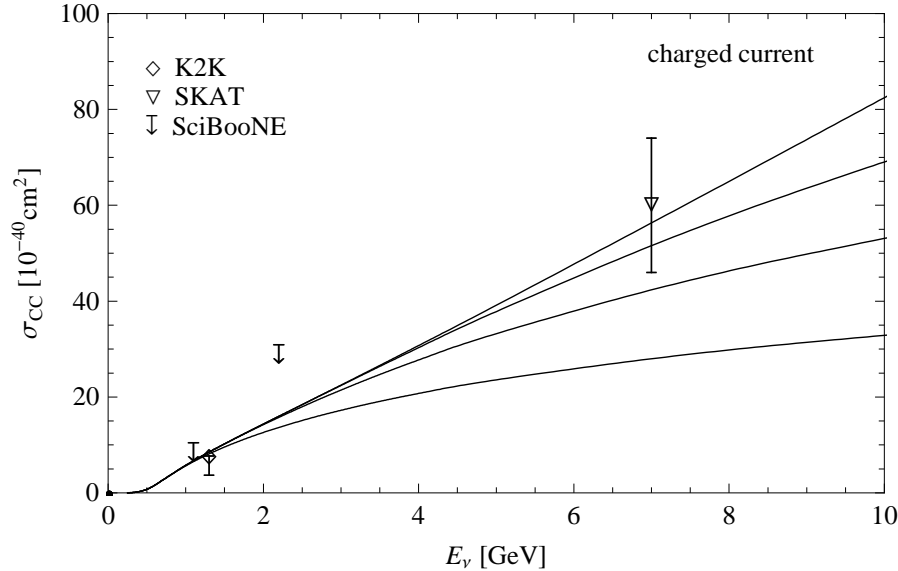}
\caption{Integrated charged current cross section with $Q^2_{max}=0.2, \; 0.5, \; 1.0 \text{ and } 4.0\unit{GeV^2}$ for $\xi=0$ (bottom to top). See \cite{Hasegawa:2005td,Grabosch:1985mt,Hiraide:2008eu} for experimental data.}
\label{fig:totalCCcuts}
\end{figure}

In this work we calculate integrated coherent cross sections for various upper values of $Q^2_{max}=0.2$, 0.5, 1.0 and $4.0\unit{GeV^2}$. The results are shown in figure \ref{fig:totalNCcuts} for neutral current and figure \ref{fig:totalCCcuts} for charged current reactions, where the effect of $Q^2_{max}$ is important.


For comparison we also included experimental results from several groups \cite{Faissner:1983ng,Raaf:2005up,Isiksal:1984vh,Grabosch:1985mt,Hasegawa:2005td,Hiraide:2008eu}. The experiments use different targets which we must rescale to the carbon target. Since the pion-nucleus cross section is a main input in this work we must rescale the results according to

\begin{equation}
\sigma_{carbon} = \sigma_{exp} \left( \frac{A_{carbon}}{A_{exp}} \right)^\frac{2}{3}
\end{equation}
which has been established in pion-nucleus elastic scattering \cite{Ashery:1981tq}. With this rescaling the agreement at higher energies prefers a $Q^2_{max}$ between 1.0 and $4.0\unit{GeV^2}$. For the energy of the neutrinos we use an average value from the neutrino flux.

There are two ways to account for the $Q^2$-dependence:
\begin{enumerate}
\item	Whenever possible, introduce an experimental cut-off to the coherent events where our model is valid.
\item Integrate the cross sections on equations (\ref{CC}) and (\ref{NC}) with a phenomenological factor $\left( \frac{m_A^2}{Q^2 + m_A^2} \right)^2$, introduced by other authors \cite{Rein:1982pf,Belkov:1986hn}, which represents the effects of heavier vector mesons and treat $m_A$ as a parameter.
\end{enumerate}

\section{Summary}
\label{sec:Summary}
Neutrino induced coherent pion production is described adequately by the method described in reference \cite{Paschos:2005km}. It is argued again that chiral symmetry relates coherent production of pions to the pion-nucleus elastic scattering in a general way, provided that $Q^2 \approx (\text{a few})\cdot m_\pi^2$.

Our formalism and results are close but not identical to a recent article \cite{Berger:2008xs}, where the elastic pion-carbon data are now used as well, which reduces their earlier predictions.
We both use elastic pion-nucleus scattering data and the small difference arise from the handling of the experimental data and the limits of integrations.

Using pion-carbon scattering data we presented quantitative results for many cross sections. We emphasize that kinematic limits, as in equation (\ref{trange}) for $t$ and the maximum value of $Q^2$ in figures \ref{fig:e1051nxi0}-\ref{fig:totalCCcuts}, are important. An overview of the double-differential cross section is shown in figure \ref{fig:dE1cc} and \ref{fig:dE10cc} showing the main characteristics of the process. The calculation can be extended to higher energies when the value of $Q^2_{max}$ is better understood.

\end{document}